\documentclass[twocolumn]{autart}  % Comment this line out
\usepackage{color}
\usepackage{graphicx}

\usepackage{hyperref}
\usepackage{amsmath,amsfonts, amssymb}
\usepackage{multirow}
\usepackage{natbib}
\begin{document}

\begin{frontmatter}
\title{A Moving-Horizon Hybrid Stochastic Game for Secure Control of Cyber-Physical Systems\thanksref{footnoteinfo}} % Title, preferably not more 
                                                % than 10 words.

\thanks[footnoteinfo]{This material is based on research sponsored by DARPA under agreement number FA8750-12-2-0247.  The U.S. Government is authorized to reproduce and distribute reprints for Governmental purposes notwithstanding any copyright notation thereon.  The views and conclusions contained herein are those of the authors and should not be interpreted as necessarily representing the official policies or endorsements, either expressed or implied, of DARPA or the U.S. Government. This work was also supported in part by NSF CNS-1505701, CNS-1505799 grants, and the Intel-NSF Partnership for Cyber-Physical Systems Security and Privacy. This paper was not presented at any IFAC meeting. Part of the results in this work appeared at the 52nd Conference of Decision and Control, Florence, Italy, December 2013~\cite{cdc_replay} and the 53rd Conference of Decision and Control, Los Angeles, CA, USA, December 2014~\cite{game_cdc14}. Corresponding author F.~Miao. Tel. 2154216608.}

\author[Uconn]{Fei Miao}\ead{fei.miao@uconn.edu},  % Add the 
\author[NYU]{Quanyan Zhu}\ead{quanyan.zhu@nyu.edu},               % e-mail address 
\author[Duke]{Miroslav Pajic}\ead{miroslav.pajic@duke.edu},  % (ead) as shown
\author[Upenn]{George J. Pappas}\ead{pappasg@seas.upenn.edu}

\address[Uconn]{ University of Connecticut, Storrs, CT, USA}
\address[Upenn]{ University of Pennsylvania, Philadelphia, PA, USA}  % Please supply                                              
\address[NYU]{ New York University, Brooklyn, NY, USA}             % full addresses
\address[Duke]{Duke University, Durham, NC, USA}
%\address[Pappas]{ University of Pennsylvania, Philadelphia, PA, USA}

\begin{keyword}
Stochastic Game, Secure Control, Saddle-Point Equilibrium
\end{keyword}
%\maketitle

\begin{abstract}
\label{abstract}
 In this paper, we establish a zero-sum, hybrid state stochastic game model for designing defense policies for cyber-physical systems against different types of attacks. With the increasingly integrated properties of cyber-physical systems (CPS) today, security is a challenge for critical infrastructures. Though resilient control and detecting techniques for a  specific model of attack have been proposed, to analyze and design detection and defense mechanisms against multiple types of attacks for CPSs requires new system frameworks. Besides security, other requirements such as optimal control cost also need to be considered. The hybrid game model we propose contains physical states that are described by the system dynamics, and a cyber state that represents the detection mode of the system composed by a set of subsystems. A strategy means selecting a subsystem by combining one controller, one estimator and one detector among a finite set of candidate components at each state. Based on the game model, we propose a suboptimal value iteration algorithm for a finite horizon game, and prove that the algorithm results an upper bound for the value of the finite horizon game. A moving-horizon approach is also developed in order to provide a scalable and real-time computation of the switching strategies. Both algorithms aims at obtaining a saddle-point equilibrium policy for balancing the system's security overhead and control cost. 
%This approach leads to a real-time algorithm that yields a sequence of Nash equilibrium strategies which can be shown to converge. 
The paper illustrates these concepts using numerical examples, and we compare the results with previously system designs that only equipped with one type of controller. 
\end{abstract}

\end{frontmatter}

\section{Introduction}
\label{sec:intro}
Cyber-Physical Systems (CPS) feature a tight integration of embedded computation, networks, controlled physical processes, and provide the foundation of critical infrastructures such as transportation systems, smart grids, water service systems and so on~(\cite{PRKumar_cps}). However, the integration structures also result in vulnerability under malicious attacks (\cite{challenge}). Recoded incidents caused by attacks show that CPS attacks can disrupt critical infrastructures and lead to undesirable, catastrophic consequences~(\cite{scada_ex}). While cyber security tools have focused on prevention mechanisms, there are still challenges on how to leverage the ability of control systems to keep system resilient under a smart adversary. 

%The vulnerability of CPS to malicious attacks result from their interaction structures among continuous physical dynamics, discrete communications, and computation substrates. 
Detection methods for various types of attacks have been analyzed in the literature. \cite{Bullo_di} propose a framework for attacks and monitors of CPS perspectives. \cite{Mo_grid} analyze security challenges and countermeasures in smart grids. \cite{res_estimator} present resilient state estimators for systems with noise and modeling errors. \cite{GPS_spoof} analyzes spoofing attacks against cryptographically-secured Global Navigation Satellite System (GNSS) signals and detection strategies. \cite{Miao_coding_tcns16} design a coding scheme for sensor outputs to detect stealthy data injection attacks over the communication channel. %Four different jamming attack models and corresponding detection schemes are analyzed by~\cite{jam}. 

In general, attack models are used as parameters to design defense schemes. However, a specific detection approach alone is not sufficient, when the system does not have knowledge which attack will happen among various types of potential attacks. CPS are usually resource constrained systems, which prevents running all available modules at the same time. Besides security, other requirements like optimal cost need to be addressed during control systems design. Consequently, considering control and defense costs with the effects of multiple attacks, strategic methods that balance the system performance and security requirements are necessary. In this work, we consider the case that at each time instant, only one detector is active because of the limits of resources. Our approach can be generalized to more than one detector being active at every time instance.

The application of game theory to security problems has raised a lot of interest in recent years. Selected works that apply game-theoretic approaches in computer networks security and privacy problems are summarized by~\cite{gt_ns}. \cite{MartZhu_game} propose a receding-horizon dynamic Stackelberg game model for systems under correlated jamming attacks. \cite{Basar_2015} propose game-theoretic methods for robust and resilient control of CPSs. However, none of these works have considered switching policies under multiple types of attacks, with payoffs as functions of system dynamics and probabilistic detection rate.%A minimax game in the presence of faults is discussed in~\cite{minimax}. 

Building a framework that captures the hybrid system dynamics and interactions with attacks is pivotal for security analysis and design of CPS. To achieve this goal, our first step is to establish a zero-sum hybrid stochastic game model. The hybrid state of the game model contains a dynamic system state that captures the evolution of the physical processes, and discrete cyber modes that represent different security states of the CPS according to information provided by the detector. Then a suboptimal value iteration algorithm is developed for the finite horizon hybrid stochastic game. Compared with our previous game model~(\cite{cdc_replay}) that only switches between two controllers against replay attacks and needs strategy history to calculate a strategy, in this work the hybrid state stochastic game strategy calculation process does not depend on the strategy history. % In this work, we consider a hybrid state game for a system with multiple types of state estimators, controllers and detectors against various types of attacks, and the suboptimal algorithm for calculating game strategies.

We then propose a moving-horizon computation methodology to reduce the computational complexity of finding a saddle-point equilibrium for the hybrid stochastic game. This is a scalable and computationally efficient algorithm. At each stage, the system selects a window of finite length for the physical state, and computes the stationary saddle-point strategies for the associated finite stochastic game, with the game state reformulated as the joint cyber and physical states. A preliminary result of the moving-horizon algorithm appeared in the conference paper~\cite{game_cdc14}; in this journal version, we have included more detail about different types of attacks and each element of the game model, revised analysis of the moving horizon algorithm compared with the suboptimal algorithm, and added more simulation results. The cost comparison with the suboptimal algorithm shows that the real-time algorithm does not sacrifice system performance much. The contributions of this work are summarized as follows: 
\begin{enumerate}
\item We formulate a zero-sum, hybrid stochastic game framework for designing a switching policy for a system under various types of attacks. 
\item We design a suboptimal algorithm for the finite horizon hybrid stochastic game, and prove that the algorithm provides an upper bound for the optimal cost of the system. 
\item We develop a real-time algorithm to reduce the computation overhead of the game model.   
\end{enumerate}

This paper is organized as follows. We describe the system, attack models, and motivation of game-theoretic techniques for switching policies in Section~\ref{sec:replay1}. In Section~\ref{sec:game_form}, we formulate a zero-sum,  hybrid stochastic game between the system and the attacker. A suboptimal algorithm for the finite horizon game is developed in Section~\ref{sec:algorithm_finite}. The moving horizon algorithm and its computational complexity are analyzed in Section~\ref{sec:algorithm}. Section~\ref{sec:simulation} compares the complexity and system performance of the finite horizon and the receding horizon algorithms. Finally, Section~\ref{sec:concl} provides concluding remarks.

\iffalse
When the sequence of game strategies converges, the state transition probability of the game converges, and we leverage the stability analysis of Markov jump systems~(\cite{delay_mlj}) to check system stability.
 \cite{taxreplay} presents a taxonomy of replay attacks on cryptographic protocols.
that uses a moving window to select a sequence of physical state information, and computes a stationary saddle-point equilibrium strategy with the state being a joint cyber and physical state 
\fi

\section{Switched System and Attack Model}
\label{sec:replay1}
\begin{figure}[b!]
\centering
\includegraphics [width=0.38\textwidth]{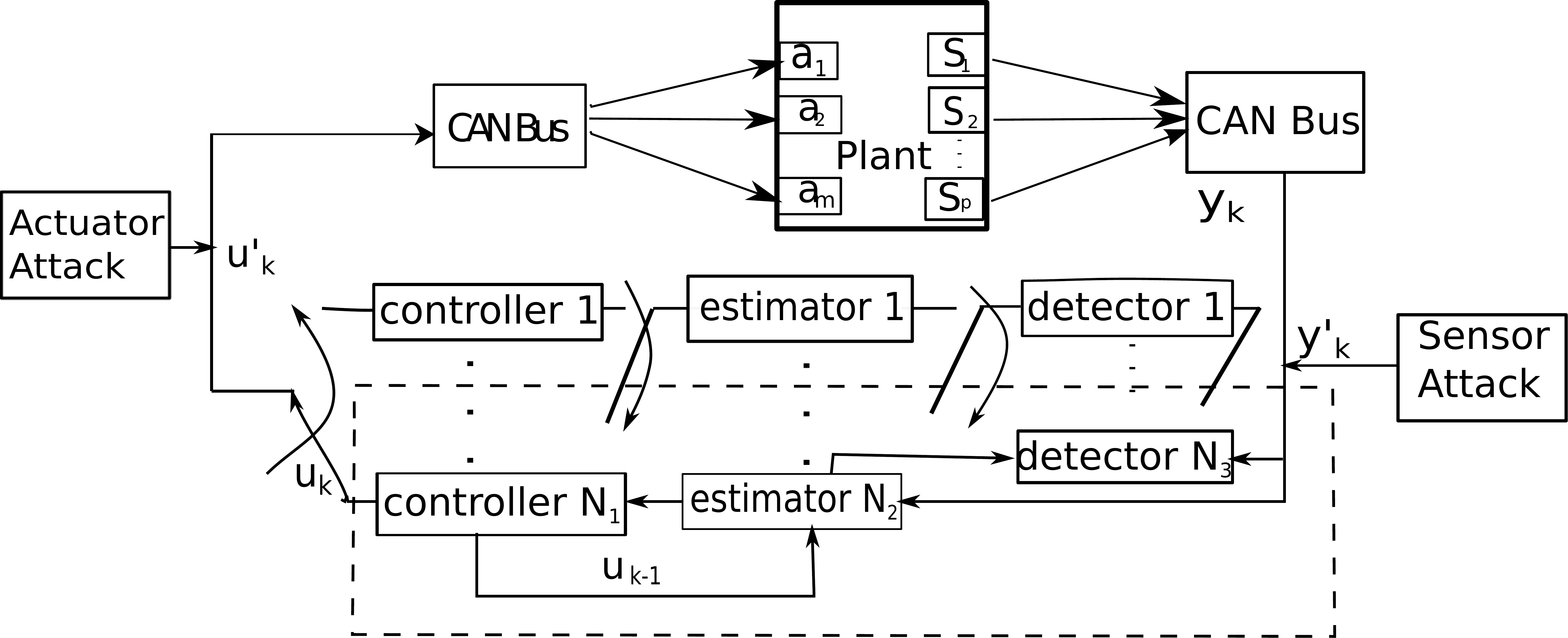}
\vspace{-10pt}
\caption{Switching system diagram, where the system is equipped with $N_1$ controllers, $N_2$ estimators and $N_3$ detectors and switches among $N$ subsystems. A subsystem (controller $N_1$, estimator $N_2$, and detector $N_3$) is chosen here.} %attacks can change sensor measurements in this example.}
\label{system}
%\vspace{-10pt}
\end{figure}
%\FM{add: fault detection filter and detector here?}
We consider the CPS security problem when both the system and attacker have limited knowledge about the opponent. The system is equipped with multiple controllers/estimators/detectors, such that each combination of these components constitute a subsystem. A subsystem has a probability to detect specific types of attacks with different control and detection costs. To balance the security overhead and the control cost under various attacks, we consider switching among subsystems (choose a model for every component) according to the system dynamics and detector information. A switched system model is shown in Figure~\ref{system}, and the model of each component is described with a concrete example in the rest of this section. It is worth noting that the set of subsystems is not restricted and can be further generalized. %We assume that the controller/estimator/detector are not compromised, and attackers can not hack code implemented in these components in this work.

\textbf{LTI plant and sensor attack model}: Consider a class of LTI plants described by: % standard state-space form:
%
%\vspace{-3pt}
\begin{align}
\begin{split}
\mathbf{x}_{k+1}= \mathbf{Ax}_{k}+\mathbf{Bu}_{k}+\mathbf{w}_{k},\quad
\mathbf{y}_{k}= \mathbf{Cx}_{k}+\mathbf{v}_{k},
\end{split}
\label{system}
\end{align}
where $\mathbf{x}_{k} \in \mathbb{R}^{n}, \mathbf{u}_{k} \in \mathbb{R}^{p}$ and $\mathbf{y}_{k} \in \mathbb{R}^{m}$ denote the discrete time state, input and output vectors respectively, and $\mathbf{w}_{k}\sim \mathcal{N}(0,\mathbf{Q})$, $\mathbf{v}_{k}\sim  \mathcal{N}(0,\mathbf{R})$ are independent and identically distributed (IID) Gaussian random noise. The initial state is $\mathbf{x}_{0}\sim  \mathcal{N}(\bar{\mathbf{x}}_{0}, \Sigma)$. Sensors or the communication between sensors and estimators are vulnerable, and attacker can change values $\mathbf{y}_k$ that sent from sensors of system~\eqref{system}, and the compromised sensor measurements are defined as $\mathbf{y}'_k$ according to the types of attacks we consider. For instance, if the attacker can inject arbitrary data $\mathbf{y}^a_k$ to sensors, $\mathbf{y}'_k=\mathbf{y}_k+\mathbf{y}^a_k$; for replay attacks, the attacker can choose the replay window size $T_2$, let $\mathbf{y}'_{k} =\mathbf{y}_{k-T_2}$ and decide whether to send the delayed plant outputs at $k$. 

\textbf{Estimators}: % Kalman filter is widely applied for noisy systems. 
The physical dynamical state of the system is provided by an estimator, for instance, attack resilient estimator~(\cite{res_estimator}), $l_1$ norm state estimator~(\cite{resest_cdc15}), fault detection filter~\cite{fd_filter}, or the widely applied Kalman filter.
When $(\mathbf{A},\mathbf{B})$ is stabilizable, $(\mathbf{A},\mathbf{C})$ is detectable, a steady state Kalman filter exists.

\textbf{Controllers}: A state feedback control law is described as $\mathbf{u}_k=L(\hat{x}_{k|k})$, where $L(\cdot)$ is a linear function, $\hat{x}_{k|k}$ is the estimated state. \cite{replay} increase the detection rate by adding an IID Gaussian signal $\Delta \mathbf{u}_{k} \sim \mathcal{N}(0, \mathcal{L})$ to $\mathbf{u}_k^*$ to an optimal LQG controller as $\mathbf{u}_{k} = \mathbf{u}^{*}_{k} + \Delta \mathbf{u}_{k} \label{nlqg}$, and increase the control cost. Then always applying the non-optimal controller for detecting a replay attack is not cost optimal, especially when there is no replay at all during a long time.

\textbf{Detectors}: We assume that every detector of the subsystem provides a detection rate for a specific type of attack, and a system is equipped with several detectors in order to deal with multiple types of attacks. Researchers have designed probabilistic detectors with respect to different attacks. For instance, \cite{fd_filter} design a fault detection filter, including a residual estimator and a threshold and a decision logic unit. Hypothesis testing strategies such as maximum likelihood (MLE), maximum a posteriori (MAP), and minimum mean square error (MMSE) account for GPS spoofing attack is presented by~\cite{GPS_spoof}. %By testing the statistical profile of a spoofing attack with methods  the probability and cost of detecting one spoofing attack are evaluated. 

%However, each method of hypothesis testing has the largest detection rate with respect to a different type of spoofing attack with some cost. We need to decide which detection strategy to use in order to secure the system with some probability and not cost too much. 

\textbf{Cyber state -- discrete modes of the system}:
We denote the modes of a vulnerable system as three constants $S=\{\delta_1,\delta_2, \delta_3\}$. State $\delta_{1}=safe$ describes that the system has already successfully detected an attack; $\delta_{2}= no~detection$ specifies that the alarm is not triggered; finally, the system enters state $\delta_{3}= false~alarm~trigger$ when the alarm is triggered while no attack has yet occurred. The mode depends on the probability detection rate. We assume that once the alarm is triggered, the system will stop the execution and decide whether to react to occurred attacks or it is a false alarm. %When the system is hijacked, the estimator, detector and controller are fed with false data, until an alarm is triggered and the system reacts to the attack.

\section{A Hybrid Stochastic Game Model}
\label{sec:game_form}
To obtain a switching policy that minimizes the expected real-time worst case payoff for the given subsystems, 
we formulate a zero-sum, hybrid stochastic game between the system and the attacker. System dynamics knowledge are combined with the game definition, and the quantitative process for the game parameters will be introduced in this section. We assume that one game stage $k$ is also one time step of the physical system. The total stage number is $K$. The hybrid game state space $(X_{[k-T,k]}\times S)$ contains information about both the system dynamics $\mathbf{x}_k$ and the discrete modes $\delta_l, l=1,2,3$. Here, $T$ is the window size of system dynamics needed to keep the state transition between stages $k$ and $(k+1)$ Markov. The joint state includes information we need to compute the game strategy at the current stage. This is the main difference compared with the previous work~(\cite{cdc_replay}), while the latter is not Markov since it needs to consider all the possible histories of strategies for deciding the physical dynamics and getting a strategy. At each stage $k \in \{T,\cdots, K+T\}$, parameters include the action space for the attacker (system) $A_{t}$ ($A_{s}$), the state transition probability matrix $\mathbb{P}_{k}$, and the immediate payoff matrix $r_{k}$. The solution set of the game is mixed strategies $\mathbf{F}_{k}$ for the attacker, and $\mathbf{G}_{k}$ for the system. Formally, the game is defined as a sequence of tuples:
 $\{(X_{[k-T,k]} \times S),A_{t},A_{s}, \mathbf{F}_{k},\mathbf{G}_{k}, P, r\}$.

\iffalse
\begin{table*}
\centering
%\caption{Parameters of the hybrid stochastic game between the system and the attacker}
\begin{tabular}{|c|c|}
  \hline
   $s_{kl}=(x_{[k-T,k]}, \delta_l)$& Joint game state: sequence of physical dynamics, and piecewise constant mode \\ \hline
   $A_{t}$ & Attacker's action space \\ \hline
   $A_{s}$ & System's action space \\ \hline  
   $\mathbf{f}_k(s_{kl})$ & Strategy of the attacker at stage $k$, 
                                           state $s_{kl}, l=1,2,3$  \\ \hline     
      $\mathbf{g}_k(s_{kl})$ & Strategy of the system at stage $k$, 
                                            state $s_{kl}, l=1,2,3$  \\ \hline     
    ${P}(s_{(k+1)h}|s_{kl})$ & Probability that system transits from state $s_{kl}$ 
                                              at stage $k$ to state $s_{(k+1)l}$ at stage $k+1$\\ \hline
    ${r}(s_{kl})$ & Immediate payoff matrix at stage $k$ \\ 
   \hline
\end{tabular}
\centering
%\captionsetup{justification=centerlast}
\caption{Parameters of the hybrid stochastic game between the system and the attacker}
\label{game_parameter}
\end{table*} 
\fi

\begin{figure}[t!]
%\vspace{-5pt}
\centering
\includegraphics [width=0.32\textwidth]{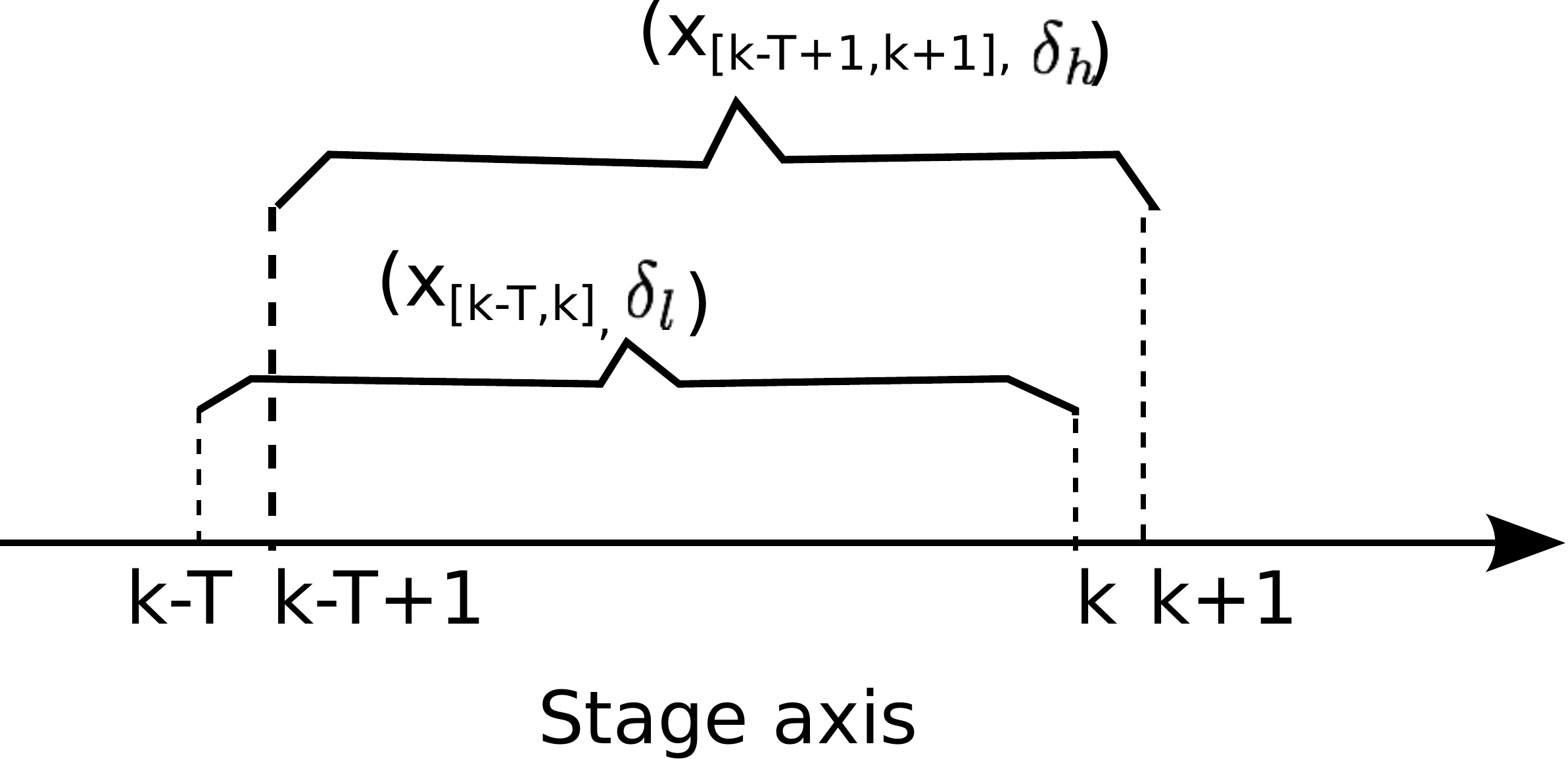}
\vspace{-8pt}
\caption{Joint state transition of the hybrid stochastic game when moving the horizon of game state one step ahead. When the state transits from stage $k$ to $k+1$, we slice the window of the sequence of physical dynamics one step ahead, add $x_{k+1}$ and remove $x_{k-T}$,  thus $x_{[k-T,k]} \to x_{[k-T+1,k+1]}$. The piecewise constant modes $\delta_l$, $\delta_h$ describe the cyber states provided by the detector at stage $k$, respectively.}
\label{sg}
\vspace{-5pt}
\end{figure}

\textbf{Game State Space}: The joint state of the system at stage $k$ is described by the pair $s_{kl}=(x_{[k-T,k]}, \delta_l)$, where
\centerline{$
x_{[k-T,k]}=(x_{k-T}, x_{k-T+1}, \cdots, x_k ) \in X_{[k-T,k]}$} is the discrete-time dynamics of the physical process provided to the system--the state estimations $\hat{x}_{k-T},\cdots, \hat{x}_k$, $\delta_l \in S=\{\delta_1,\delta_2, \delta_3\}$ denote the cyber state of the system. We assume that once the game reach $\delta_1$, the system wins and will not enter other modes till next game, i.e., $\delta_1$ is an absorbing state. The moving-horizon transition of the joint states on stage axis is shown as Figure~\ref{sg}. The window size of system dynamics $T$ keeps the state transition between time $k$ and $k+1$ Markov. For instance, if the detector of the system requires system dynamics $\hat{x}_{[k-T_1,k]}$, and we consider sensor data injection attacks and replay attacks with replay windows less than $T_2$ steps, then $T=max\{T_1, T_2\}$. 
%With the system dynamics $\hat{x}_{[k-T, k]}$ denoted as $x_{[k-T,k]}$ for game stage $k$,  information needed to define the following action space of two players, the payoff and state transition probability is included. 

%i.e., the detector needs information for T steps to decide whether the alarm should be triggered. 
%\FM{In replay attack, we can say: once alarm is triggered, the system can stop the execution and check whether attack occurred, is this true for other attacks? Can the system distinguish between successfully detection and false alarm trigger?}

%For simplicity, we omit the subscript k and just write state at every time k as $s_i,  i=1,2,3$.
\textbf{Attacker's Action Space}: We assume that the system is vulnerable to different attack models described by the action space $A_{t}$, where 
\\\centerline{$
A_{t}=\{a_{1}(x_{[k-T,k]}), a_{2}(x_{[k-T,k]}), \cdots, a_{M}(x_{[k-T,k]})\}
$}
is the attacker's action space at stage $k$, and $a_{1}$ means no attack. Here we only consider discretized action space of the attacker for computational efficiency. For the LTI system dynamics considered in this work, the distance of a continuous point to its nearest discrete point in action space is bounded. With bounded error of the dynamics by discretized continuous action space, the quality of game solutions under different conditions is analyzed by work~\cite{disaction}. 

The actions can describe both multiple types of attacks and the same type attack with different values. For instance, when considering only sensor data injection attacks with different norms of injection value, we will denote $a_i (x_{[k-T,k]}), i=2, 3,\dots$ as changing the sensor value from $\mathbf{y}_k=\mathbf{Cx}_k+\mathbf{v}_k$ to $\mathbf{y}'_k= \mathbf{y}_k+\mathbf{y}_{k,i}^a$, where any injection $\mathbf{y}_k^a$ is classified as $a_i (x_{[k-T,k]}), i=inf\{i:\ \|\mathbf{y}_k^a-\mathbf{y}_{k,i}^a\|_2\}$ in attacker's action space.
Similarly, for replay attack only, the action space is discretized as changing sensor values from $\mathbf{y}_k=\mathbf{Cx}_k+\mathbf{v}_k$ to $\mathbf{y}'_k= \mathbf{y}_{k-T_i}$ for action index $a_i (x_{[k-T,k]})$, where any replay time length $T_a$ is classified as $a_i (x_{[k-T,k]}), i=inf\{i:|T_a - T_i|\}$. Considering multiple types of attacks, we assume that the system is valnerable under $m_a$ types of attacks, and attack type $A_i$ is corresponding to $M_{a,i}$ discretized actions in the action space, then there are $\sum_{i=1}^{m_a} M_{a,i}+1$ actions in total within the attacker's action space $A_t$.

% with discretized, bounded norm $\|\mathbf{y}_{k,i}^a\|_2 \leqslant b$, since the attacker has limited energy for every data injection. This means any injection data that satisfies $\|\mathbf{y}_{k}^a\|_2 \leqslant b$ is considered as $inf\{i:\ \|\mathbf{y}_k^a-\mathbf{y}_{k,i}^a\|_2\}$ in attacker's action space. 

%For example, 
%when considering replay attacks and false data injection attacks, we take $a_{2k}$ as~\eqref{replay_y} for a given replay window size, and $a_{3k}$ as~\eqref{attackmodel} for a given data injection range. 

\iffalse by any given controller/estimator/detector combination of the system \fi
%; here, $y_{k}$ is the real sensor value and $\mathbf{y}_{k-t_{i}}$ denotes any replay sensor value in the strategy set. %\FM{here we assume during time $k \in \{1,...,K\}$ the replay window size $T$ does not change for simulation.}
%\item
%\FM{rewrite the action space definition}

\textbf{System's Action Space}: The system's action space at stage $k$ is defined as
\\\centerline{$
A_{s}=\{u_{1}(x_{[k-T,k]}), u_{2}(x_{[k-T,k]}),\cdots, u_{N}(x_{[k-T,k]})\},
$}  %= \{\mathbf{u}^{*}_{k}, \mathbf{u}^{*}_{k} + \Delta \mathbf{u}_{k}\}$
where $u_{j}$ is the index for the $j$th subsystem. We assume that the $N$ subsystems (a model for each component in Figure~\ref{system}) are determined priorly. For example, a subsystem can be the plant with a given optimal LQG controller, a Kalman filter and a $\chi^2$ detector. A subsystem can also be the plant with an optimal LQG controller, a resilient state estimator~\cite{res_estimator} and its corresponding estimation residual checking component. We assume that the attacker's action space is defined, with corresponding system's action or a subsystem that the detection rate is greater than $0$. A switched system does not ensure performance under the attack outside the action space of the game.

\textbf{Mixed Strategy}: Let $f^{i}_{k}(s_{kl})$ ($g^{j}_{k}(s_{kl})$) be the probability that the attacker (system) chooses action $a_{i}(x_{[k-T,k}) \in A_{t}$ ($u_{j}(x_{[k-T,k}) \in A_{s}$) at state $s_{kl}\in (X_{[k-T,k]}\times S)$. Define $\mathbf{F}_{k}$ and $\mathbf{G}_{k}$ as the mixed strategy sets of the attacker and the system for stage $k$:
$\mathbf{F}_{k} :=\{\mathbf{f}_{k}= [\mathbf{f}_{k}(s_{k1}), \mathbf{f}_{k}(s_{k2}), \mathbf{f}_{k}(s_{k3})]
|f_{k}^{i}(s_{kl})\geq 0, \mathbf{f}_k \in [0,1]^{M\times 3},
\sum \limits_{a_{ik} \in A_{tk}}f_{k}^{i}(s_{kl}) = 1,\mathbf{f}_{k}(s_{kl})\in \mathbb{R}^{M}, \forall s_{kl} \in(X_{[k-T,k]}\times S)\},$
$\mathbf{G}_{k}:=\{\mathbf{g}_{k}= [\mathbf{g}_{k}(s_{k1}), \mathbf{g}_{k}(s_{k2}), \mathbf{g}_{k}(s_{k3})]|$
$g_{k}^{j}(s_{kl})\geq 0, \mathbf{g}_k \in [0,1]^{N \times 3}, %&\forall u_{jk} \in A_{sk},%k \in \{1,...,K\},\\
\sum \limits_{u_{jk} \in A_{sk}}g_{k}^{j}(s_{kl}) = 1, \mathbf{g}_{k}(s_{kl}) \in \mathbb{R}^{N}, \forall s_{kl} \in (X_{[k-T,k]}\times S)\}. $ Note that $\mathbf{x}_{[k-T,k]}$ provides exogenous information for the strategy $\mathbf{f}_k (\mathbf{g}_k)$, since for every $l$, $\mathbf{f}_{k}(s_{kl}) (\mathbf{g}_{k}(s_{kl}))$ is the strategy at mode $\delta_l$ for the same $\mathbf{x}_{[k-T,k]}$ at stage $k$. Hence, $\mathbf{g}_k$ and $\mathbf{f}_k$ are finite dimensional vectors, that the stationary strategy chosen by each player at stage $k$ depends on the cyber state. %Mixed strategy set $\mathbf{F}_k$ also include the case that the attacker only implement one specific type of attack in the action space at time instance $k$, since we do not have know the strategy of the attacker, we are able to consider all possible combinations of attacks by exploiting mixed strategies. 

%\textbf{}:
%\label{dynamicgame}
%With all the above definition, %for any strategy history $h_{k}$ (a sequence of switching policy), 

\textbf{System and Subsystem Dynamics under game framework}: Given the subsystem and attack models in Section~\ref{sec:replay1} and the game definition,  
%we can transform it to a new system dynamic model decided by both players' actions. 
we show the dynamics at stage $k$ given an action pair $(a_{i}(x_{[k-T,k}),u_{j}(x_{[k-T,k}))$ (assume initial $\mathbf{\hat{x}}_{1|0}=\bar{\mathbf{x}}_{0}$, $\mathbf{x}_{1}=\mathbf{x}_{0}$). Each action pair $(a_{i}(x_{[k-T,k]}),u_{j}(x_{[k-T,k]}))$ defines the corresponding system dynamics at $k$. For instance, when we focus on sensor attacks (like replay or false data injection), let $\mathbf{\gamma}_{k}(a_{i}(x_{[k-T,k]}), u_{j}(x_{[k-T,k]}))$ be the control input with $(a_{i}(x_{[k-T,k]}),u_{j}(x_{[k-T,k]}))$, a subsystem $u_{j}(x_{[k-T,k]})$ with a Kalman filter, an optimal LQG controller has the following dynamics (we denote $(a_{i}(x_{[k-T,k]}),u_{j}(x_{[k-T,k]}))$ as $(a_{ik}, u_{jk})$ for convenience): 
\begin{align}
\begin{split}
&\mathbf{x}_{k}=\mathbf{Ax}_{k-1}+ \mathbf{Bu}_{k-1}+\mathbf{w}_{k-1},\\
& \mathbf{y}_{k}=\begin{cases}a_{1k} = \mathbf{Cx}_k+\mathbf{v}_{k},\ \text{without attack}\\
a_{ik}, i=2,\cdots, M, \ \ \text{with attack,} \end{cases}\\
&\hat{\mathbf{x}}_{k|k-1}= \mathbf{A\hat{x}}_{k-1|k-1}+\mathbf{Bu}_{k-1},\\
%&\mathbf{z}_{k+1}(h_{k},a_{ik},u_{jk})=a_{ik}(h_{k}) - \mathbf{C\hat{x}}_{k+1|k}(h_{k},a_{ik},u_{jk}),\\
&\hat{\mathbf{x}}_{k|k}(a_{ik}) =\hat{\mathbf{x}}_{k|k-1}+ \mathbf{K}(a_{ik} - \mathbf{C\hat{x}}_{k|k-1}),\\
%\end{split}
&\mathbf{\hat{x}}_{k+1|k}(a_{ik},u_{jk})=\mathbf{A\hat{x}}_{k|k}(a_{ik})+\mathbf{B\gamma}_{k}(a_{ik},u_{jk}),\\
& \mathbf{\gamma}_{k}(a_{ik}, u_{jk}) = \mathbf{L\hat{x}}_{k|k} (a_{ik}),\\%+\Delta \mathbf{u}_{k},\\
&\mathbf{z}_{k+1}(a_{ik},u_{jk})=a_{ik} - \mathbf{C\hat{x}}_{k+1|k}(a_{ik},u_{jk}).
\label{dynamicgame}
\end{split}
\end{align}
\textbf{State Transition Probability}: Given a set of subsystem models, define the state transition probability $P$ as a function of the state of the game and both players' actions $P:\ (X_{[k-T,k]}\times S) \times A_{t} \times A_{s}\to [0, 1],$
where
\\\centerline{$
P(s_{(k+1)h}|s_{kl},a_{ik}, u_{jk}), h=1,2,3
$}
%\end{align*}
is the probability that system transits from state $s_{kl}$ to state $s_{(k+1)h}$ at stage $k+1$, given both players' action $(a_{ik},u_{jk})$ at stage $k$. Given the current game state $s_{kl}=(x_{[k-T,k]}, \delta_l)$ and an action pair $(a_{ik},u_{jk})$, the dynamics of the system at stage $k+1$ is described as $x_{[k-T+1,k+1]}$ for all possible cyber modes $\delta_h \in S$, hence the dimension of state transition probability $P(s_{(k+1)h}|s_{kl},a_{ik}, u_{jk})$ is determined by the number of cyber modes of the game. We denote $P(s_{(k+1)h}|s_{kl}, a_{ik}, u_{jk})$ as $P^{ij}(s_{(k+1)h}|s_{kl})$ for short.
%and $\tilde{P}^{ij}(s_{(k+1)h}|s_{kl})$ is the entry at the $i$-th row and $j$-the column  of the state transition matrix $\tilde{P}(s_{(k+1)h}|s_{kl})$ of the game at hybrid state $s_{kl}$.
 As a state transition probability, this function should also satisfy
%\begin{align*}
\\\centerline{$\sum_{\delta_h \in S} {P}^{ij}(s_{(k+1)h}|s_{kl}) = 1,\quad \forall (a_{ik},u_{jk}) \in A_{t} \times A_{s},$}
\\\centerline{$s_{(k+1)h} \in (X_{[k-T+1,k+1]}\times S), s_{kl} \in(X_{[k-T,k]}\times S).$}
%\end{align*}
The transition probability is provided by intrusion detectors of the subsystem. 
%For computational efficiency, we assume that every element of the state transition matrix is a convex function of the system dynamics $x_{[k-T,k]}$ or can be convexified with bounded error. 
%For example, if a $\chi^{2}$ detector is the detector component of subsystem $u_{j}$, we apply~\eqref{alarm} to decide the state transition probability.

\textbf{Immediate Payoff Function}: The immediate payoff matrix at stage $k$ is a $\mathbb{R}^{M\times N}$ matrix for given game state and every action pair $(a_{ik}, u_{jk})$. We define the immediate payoff function as a continuous, convex function of the hybrid game state and the actions of both players
\\\centerline{$r: (X_{[k-T,k]}\times S) \times A_{t} \times A_{s} \to \mathbb{R}^{M \times N},$}
where $r(s_{kl}, a_{ik}, u_{jk}) \geqslant 0$ is the payoff at joint state $s_{kl}$ given action pair $(a_{ik}, u_{jk})$. For definition convenience, we denote ${r}(s_{kl}, a_{ik}, u_{jk})$ as ${r}^{ij}(s_{kl})$ for short, since it is the element on the $i$-th row and $j$-th column of the payoff matrix ${r}(s_{kl})$. It is a zero-sum game between the system and the attacker, and we assume the system is the minimizer and the attacker is the maximizer, hence the payoff function for the attacker and the system is defined as
\centerline{$
{r}^{ij}(s_{kl})={r}_t^{ij}(s_{kl})=-{r}_s^{ij}(s_{kl}).
$}
For instance, when the linear quadratic cost is a metric of system performance, let $\gamma_{k}(a_{ik}, u_{jk})$ be the control input given action pair $(a_{ik}, u_{jk})$, then the payoff function is defined as
\begin{align}
\begin{split}
{r}^{ij} (s_{k1}) =&\mathbb{E}[\mathbf{\hat{x}}^{T}_{k}]\mathbf{W}\mathbb{E}[\mathbf{\hat{x}}_{k}]+\mathbb{E}[\mathbf{\gamma}^{T}_{k}(a_{1k},u_{jk})]\mathbf{U}\mathbb{E}[\mathbf{\gamma}_{k}(a_{1k},u_{jk})],\\
{r}^{ij} (s_{k2}) =&\mathbb{E}[\mathbf{\hat{x}}^{T}_{k}]\mathbf{W}\mathbb{E}[\mathbf{\hat{x}}_{k}]+\mathbb{E}[\mathbf{\gamma}^{T}_{k}(a_{ik},u_{jk})]\mathbf{U}\mathbb{E}[\mathbf{\gamma}_{k}(a_{ik}, u_{jk})],\\
{r}^{ij} (s_{k3}) =& p_f,
\end{split}
\label{payoff}
\end{align}
where $p_f$ is the false alarm trigger penalty, the cost that the system needs to stop execution, check the reason of an alarm, and restart later; $\mathbf{x}_{k}$ is the physical state under the game framework. At mode $\delta_{1}$ the system wins, so the payoff is a normal system payoff with correct sensor data. The larger $p_f$ is, the less probable it is for the system to choose a strategy to transit to state $s_{k3}$.

\textbf{System dynamics update with strategies at stage k}:
 Let $p(s_{kl})$ be the probability system is at state $s_{kl}$ at stage $k$. The initial state distribution $p(s_{1l})$ is given. With  a strategy $\mathbf{f}_{k},\mathbf{g}_{k}$, the attacker and the system randomly sample an action pair $(a_{ik}, u_{jk})$ according to the probability distribution. Then, the control input and sensor value for calculating expectation cost are: 
%\begin{align*}
\centerline{$
\mathbf{u}_{k}=\sum\limits_{j=1}^{N}\sum\limits_{i=1}^{M} \sum\limits_{l=1}^{3}p(s_{kl})f_{k}^{i}(s_{kl})g^{j}_{k}(s_{kl})\mathbf{\gamma}_{k}(a_{ik},u_{jk}),
$}
$\text{ }\quad\quad\mathbf{y}_{k} =\sum\limits_{i=1}^{M}\sum\limits_{l=1}^{3} p(s_{kl})f_{k}^{i}(s_{kl}) a_{ik}.$
%\end{align*}
\\The probability that system is at state $s_{(k+1)h}$ for $k+1$ is:
\\\centerline{$
%\begin{align*}
p(s_{(k+1)h})= \sum\limits_{l=1}^{3}p(s_{kl})[\mathbf{f}_{k}(s_{kl})]^{T}{P}_{k}(s_{(k+1)h}|s_{kl})\mathbf{g}_{k}(s_{kl}). 
%\end{align*}
$}

\iffalse
\begin{align*}
\mathbf{F}_{k} :=\{&\mathbf{f}_{k}= [\mathbf{f}_{k}(s_{k1}), \mathbf{f}_{k}(s_{k2}), \mathbf{f}_{k}(s_{k3})]
|f_{k}^{i}(s_{kl})\geq 0,\\& \mathbf{f}_k \in [0,1]^{M\times 3} %&\forall a_{ik} \in A_{tk},  %k\in \{1,...,K\},\\
\sum \limits_{a_{ik} \in A_{tk}}f_{k}^{i}(s_{kl}) = 1,\mathbf{f}_{k}(s_{kl})\in \mathbb{R}^{M},\\&\forall s_{kl} \in(X_{[k-T,k]}\times S)\},\\
\mathbf{G}_{k}:=\{&\mathbf{g}_{k}= [\mathbf{g}_{k}(s_{k1}), \mathbf{g}_{k}(s_{k2}), \mathbf{g}_{k}(s_{k3})]|
g_{k}^{j}(s_{kl})\geq 0,\\& \mathbf{g}_k \in [0,1]^{N \times 3}, %&\forall u_{jk} \in A_{sk},%k \in \{1,...,K\},\\
\sum \limits_{u_{jk} \in A_{sk}}g_{k}^{j}(s_{kl}) = 1, \mathbf{g}_{k}(s_{kl}) \in \mathbb{R}^{N},\\ &\forall s_{kl} \in (X_{[k-T,k]}\times S)\}. 
\end{align*} 
\fi

\section{Existence of An Optimal Strategy and Suboptimal Algorithm for A Finite Game}
\label{sec:algorithm_finite}

Based on the game formulation, in this section we discuss the existence of an optimal solution for the finite form of the hybrid stochastic game, and present an algorithm to compute a suboptimal system strategy.%\PA{what do you mean by ``approximation computing algorithm''}
 
\subsection{Existence of the System's Optimal Strategy}

%\iffalse 
We define the concatenation of strategies for $K$-stage game of each player ($\mathbf{f}$ for attacker and $\mathbf{g}$ for system) as $\mathbf{f}=\mathbf{f}_1\cdots \mathbf{f}_K,\quad\mathbf{f}_{k} \in \mathbf{F}_{k},\quad\mathbf{f}\in \mathbf{F},$
$\mathbf{g}=\mathbf{g}_{1}\cdots\mathbf{g}_K,\quad\mathbf{g}_{k} \in \mathbf{G}_{k},\quad\mathbf{g}\in\mathbf{G}$,
$k=1, 2,\dots,K$.
\begin{defn}
Let the random variable $\zeta_{k}$ describe the discrete state of the hybrid game at stage $k$, we define the conditional expected total payoff till $\tilde{K}$ for any $\mathbf{f},\mathbf{g}$ as%\begin{align*}
\\$\text{ }\quad R_{\tilde{K}}(s, \mathbf{f}, \mathbf{g})$
\\\centerline{$
=\sum \limits^{\tilde{K}}_{k=1}\sum_{l=1}^{3} p(\zeta_{k}=\delta_{l}|\zeta_{1} = s)[\mathbf{f}_{k}(s_{kl})]^{T}\tilde{r}_{k}(s_{kl})\mathbf{g}_{k}(s_{kl}),
$}
where $p(\zeta_{k}=\delta_{l}|\zeta_{1} = s)$ is the probability that the discrete state of the hybrid game is $\delta_{l}$ at stage $k$ given its initial discrete state $\zeta_{1}=s$.
\label{R_K}
\end{defn}
Since the immediate payoff of each stage satisfies that $0 \leq \tilde{r}_{k}^{ij}(s_{kl})< \infty,\ \text{for all}\ k,i,j,$
we have that $R_{\tilde{K}}(s, \mathbf{f}, \mathbf{g})$ is a nonnegative real-valued, nondecreasing function with $\tilde{K}$. 
%\PA{when you cut an old sentence to save space, make sure that the new one makes sense; for example you should have said - from the immediate payoff definition we have that ...' or ``since...''}  
Furthermore, for finite $K$%a finite stage game
%\PA{shouldn't it be ``for a finite game'' or ``for a finite stage game''}
%\vspace{-4pt}
\begin{align}
%\sup_{\mathbf{f} \in F^{K}}R_{K}(s,\mathbf{f}, \mathbf{g}^{*}) < \infty,s \in S.
%\sup_{\mathbf{f}}
R_{K}(s,\mathbf{f}, \mathbf{g}) < \infty, \forall s \in S, \mathbf{f}\in \mathbf{F}, \mathbf{g} \in \mathbf{G}.
\label{fR}
\end{align}
Similarly as the definition of value and optimal strategy for a zero-sum, finite discrete state, finite stage stochastic game, we define the value and optimal strategy for the hybrid state stochastic game defined in this work as the following.
\begin{defn}
A two-person zero-sum $K$-stage stochastic game is said to have a value vector $v^{*}_{K}$ if $v^{*}_{K,s}=\underbar{v}_{K,s}=\bar{v}_{K,s},$ for any initial cyber state $s\in S$, where
\\\centerline{$
\underbar{v}_{K,s}= \sup_{\mathbf{f}\in\mathbf{F}}\inf_{\mathbf{g}\in\mathbf{G}}R_{K}(s,\mathbf{f}, \mathbf{g}),
$}
\centerline{$
\bar{v}_{K,s}=\inf_{\mathbf{g}\in\mathbf{G}}\sup_{\mathbf{f}\in\mathbf{F}}R_{K}(s,\mathbf{f},\mathbf{g}).
$}
For the finite value $K$-stage stochastic game, strategies $\mathbf{g}^{*}$ and $\mathbf{f}^{*}$ are called optimal at the saddle-point equilibrium for player two (the system) and player one (the attacker), respectively, if for all $s\in S$,
\centerline{$
v^*_{K,s} = \inf\limits_{\mathbf{g}\in \mathbf{G}}R_{K}(s, \mathbf{f}^{*}, \mathbf{g}),\quad v^*_{K,s} = \sup\limits_{\mathbf{f}\in \mathbf{F}}R_{K}(s, \mathbf{f}, \mathbf{g}^{*}).
$}
\end{defn}
%The existence conditions of the value and optimal strategies for a general finite horizon, finite state, zero-sum stochastic game are. 
The game defined in this paper has finite action spaces, finite strategy space, finite discrete cyber modes and satisfies~\eqref{fR} with bounded total payoff in finite horizon. Therefore, there exists the value of the considered  game and an saddle-point equilibrium or optimal strategy for the system shown in~\cite{dgt_Basar}.
\subsection{Suboptimal algorithm for the finite game}
Existing value iterative algorithms or dynamic programming algorithms for finite stochastic games cannot be used to solve the finite hybrid stochastic game defined in this work, since the discrete time dynamics $x_{[k-T,k]}$ of the game at stage $k$ depends on that of the stage $k-1$, which is only available in the future algorithm iterations. Hence, we design a suboptimal algorithm based on the value iteration method for a finite horizon, finite discrete state stochastic game~(\cite{plangame}) and robust game techniques~(\cite{RGT}). The value iteration algorithm for a finite horizon, discrete state stochastic game (with fixed payoff $r$ and state transition probability $P$ at every stage) works in the way that if a player knew how to play in the game optimally from the next stage on, then, at the current stage, he would play with such strategies. The value of $K$-stage game is finally provided by the last step of iteration.

For a multi-stage game, to calculate the game value, we define the auxiliary matrix at stage $k$ for every cyber state $\delta_l$ with system dynamics $x_{[k-T,k]}$ as $Q(s_{kl}) \in \mathbb{Q}_{k} \subset \mathbb{R}^{M \times N}$, and each element of $Q(s_{kl})$ for action pair $(a_{ik}, u_{jk})$ is defined as
\begin{align}
\begin{split}
&Q^{ij}(s_{kl})\\
=&r^{ij}(s_{kl})
                         +\sum_{\delta_h\in S} {P}^{ij}(s_{(k+1)h} |s_{kl})\cdot {v}_{k+1}(s_{(k+1)h}), 
\end{split}
\label{Q_k}
\end{align} 
where ${v}_{k+1}(s_{(k+1)h})$ is the game value from stage $k+1$, state $s_{(k+1)h}$ (with cyber mode $\delta_h$) to the final stage $K$. For the final stage $K$, we define $Q(s_{Kl})=r(s_{Kl})$. We define a one-shot game at stage $k$ as a finite action space, zero-sum game between the system and the attacker with payoff matrix $Q(s_{kl})$, i.e., $Q^{ij}(s_{kl})$ is the payoff for action pair $(a_{ik},u_{jk})$ of stage $k$. In each one-shot game, the system only consider a strategy $f_k(s_{kl})$ to minimize the worst case payoff caused by the attacker according to matrix $Q(s_{kl})$. Here $Q(s_{kl})$ is defined based on the the system dynamics and the state transition probability provided by the detector. An alternative algorithm with unknown transition matrix or payoffs will be our future work.

Similarly as the value iteration algorithm for a discrete state stochastic game~(\cite{plangame}), Algorithm~\ref{finite} of the finite hybrid state stochastic game starts from the last stage, then gets the optimal one-stage strategy and the upper bound of game value at each stage. By calculating values of all stages until backwards to the first stage, Algorithm~\ref{finite} returns an upper bound for the value of the total payoff in $K$-stages. 

To estimate the values at each step, we consider the immediate payoff $r(s_{kl})$, the state transition probability ${P}(s_{(k+1)h}|s_{kl})$ and the game value estimated at the previous step uncertain parameters for the one shot robust game (\cite{RGT}). Then approximate each iteration value as the value of the robust one shot zero sum game. 
Algorithm~\ref{finite} provides an upper bound for the game value and the corresponding suboptimal strategy for the system. The idea is to solve a robust game at each iteration step -- i.e., minimize the worst-case caused by extreme points of the set of auxiliary matrix $\mathbb{Q}_k$ defined for all possible dynamics $x_{[k-T,k]}$.

To quantify the boundary of the set of auxiliary matrix $\mathbb{Q}_k$ %($r_{k}$, $\mathbb{P}_{k}$), 
we need the expected values of system dynamics $\mathbf{x}_{k}, \mathbf{u}_{k}$, $\mathbf{y}_{k}, k=1,\cdots,K$ defined in equations~\eqref{dynamicgame}, which is determined by the strategies from stage $1$ till stage $k$. 
We first analyze the uncertain sets of the immediate payoff function at stage $k$, and the extreme points for the uncertain set $\mathbb{Q}_k$ depend on pure strategies . Let $\mathbf{f}^p_{k-1}$, $\mathbf{g}^p_{k-1}$ be the concatenation of previous pure strategies of the attacker and the system till stage $k \geqslant 2$, respectively, where
\\\centerline{
$\mathbf{f}^{p}_{k-1}=\mathbf{f}^{p}_{1}\cdots\mathbf{f}^{p}_{k-1},\quad  \mathbf{g}^{p}_{k-1}=\mathbf{g}^{p}_{1}\cdots\mathbf{g}^{p}_{k-1}$
}
satisfies that all $\mathbf{f}^{p}_{t}(s)$ ($\mathbf{g}^{p}_{t}(s)$) for $t=1,2,\dots, k$ have only one non-zero element, i.e., the player chooses the corresponding action or the \emph{pure} strategy.

Define a pure strategy auxiliary matrix $Q^p(s_{kl}) \in \mathbb{Q}^p_{k}$ as:
\begin{align}
\begin{split}
%\vspace{-5pt}
&Q^{p}(s_{kl}) \\
=&r^{p}(s_{kl})+\sum_{\delta_h\in S} {P}^{p} ( s_{(k+1)h} |s_{kl})\cdot \bar{v}^p_{k+1}(s_{(k+1)h}),
\label{eq:Q}
\end{split}
\end{align}
%\normalsize
for stages $k=1,\dots,K-1$, and for the final stage $k=K$,
\begin{align}
Q^{p}(s_{Kl})=r^{p}(s_{Kl}).
\label{Q_p}
\end{align}
For each stage $k$, $\bar{v}^p_{k}(s_{kl})$ is defined as
% and relates to matrix games defined by $\mathbb{Q}_{(k+1)p}$.
\begin{align}
\bar{v}_{k}^{p}(s_{kl})=\max_{Q^{p}(s_{kl})\in \mathbb{Q}^p_{k}}v^*[Q^{p}(s_{kl})],%v_{k}^{s_{l}}(h_{k}^{p})&=\max_{(r^{p}_{k}(h^{p}_{k},s_{l}), P^{p}_{k}(h^{p}_{k},s_{l}))}
%v^*[r^{p}_{k}(h^{k}_{p},s_{l})+\sum_{s'\in S} P^{p}_{k}( s' |h^{p}_{k},s_{l})v_{k+1}^{s'}(h^{p}_{k+1})],
\label{pickv}
\end{align}
where $v^*$ is the function that yields the value of a zero-sum matrix game. Then the value $\bar{v}^p_{k+1}(s_{(k+1)h}) \geq 0$ to calculate the auxiliary matrix~\ref{eq:Q} is the upper bound of robust game value from stage $k+1$ till stage $K$, resulting from the iteration at stage $k+1$. This value iteration process is the key idea of the following~Algorithm~\ref{finite}. 

\begin{alg}
%\vspace{-5pt}
\textbf{: Suboptimal Algorithm for A Finite Hybrid Stochastic Game}\\
\textbf{Input}: System model parameters and game parameters.
\\\textbf{Initialization}:
             Compute the set of $\mathbb{Q}^p_k$ for every stage $k=T,\dots,T+ K$ given $\hat{\mathbf{x}}_{[0,T]}$;
              %For all $s_{l} \in S, l =1,2,3, h^{p}_{K}\in H^{p}_{K} :$ get %$4^{K-1}$ 
              %backup matrix set $\{Q_{k}^{p}(h_{k}^{p},s_{l})\}$:
             get the robust game value and corresponding strategies at stage $K$: $Q^{p}(s_{(K+T)l})=r^{p}(s_{(K+T)l})$,
                    $f^{*}(s_{(K+T)l}), g^{*}(s_{(K+T)l}), \bar{v}_{K+T}^{p}(s_{(K+T)l}) \leftarrow \pi(Q^{p}(s_{(K+T)l})).$
\\\textbf{Iteration}: For $k=(K+T-1), \cdots, T$, obtain a set of auxiliary matrices $\mathbb{Q}_{k}^{p}$ for all $\mathbf{f}^{p}_{k}$, $\mathbf{g}^p_k$, where each matrix is defined in~\eqref{eq:Q}, then calculate:
%\[Q_{k}^{p}(h^{p}_{k},s_{l}) = r^{p}_{k}(h^{p}_{k},s_{l})+\sum_{s'\in S} P^{p}_{k}( s' |h^{p}_{k},s_{l})v_{k+1}^{s'}(h^{p}_{k+1}),\]

 $f^{*}(s_{kl}), g^{*}(s_{kl}),\bar{v}_{k}^{p}(s_{kl}) \leftarrow \pi(Q^{p}(s_{kl}))$.
 
%$\mathbf{f}^*_k=[\mathbf{f}^{*}(h^{p}_{k}, s_{l}),\mathbf{f}^{*}(h^{p}_{k}, s_{2}), \mathbf{f}^{*}(h^{p}_{k}, s_{3})],$ $ \mathbf{g}^{*}_k=[\mathbf{g}(h^{p}_{k}, s_{1}),\mathbf{g}(h^{p}_{k}, s_{2}),\mathbf{g}(h^{p}_{k}, s_{3})].$ 

$\mathbf{f}^*_k=[\mathbf{f}^{*}(s_{kl}),l=1,2,3],$ $ \mathbf{g}^{*}_k=[\mathbf{g}^*(s_{kl}), l=1,2,3].$ 

\textbf{Return}:strategies $\mathbf{f}_{a}=\mathbf{f}^{*}_T\cdots\mathbf{f}^*_{K+T},$ $\mathbf{g}_{a}=\mathbf{g}^{*}_T\cdots\mathbf{g}^*_{K+T}$ and the value upper bound $\bar{v}_1^p(s_{1l}),l=1,2,3$.
\label{finite}
\end{alg}

Now consider the iteration for calculating $\bar{v}^p_{k}(s_{kl})$ from all matrix games $Q^{p}(s_{kl})\in \mathbb{Q}_{k}^p$ applying Algorithm~\ref{finite}. We define any strategy concatenations to stage $k-1$ with at most one non-pure strategy at stage $(k-1)$ as
\begin{align}
\begin{split}
\mathbf{f}^{np}_{k-1}&=\mathbf{f}^p_{k-2}\mathbf{f}_{k-1},\quad \mathbf{f}_{k-1} \in \mathbf{F}_{k-1},\\ \mathbf{g}^{np}_{k-1}&=\mathbf{g}^p_{k-2}\mathbf{g}_{k-1},\quad \mathbf{g}_{k-1} \in \mathbf{G}_{k-1},
\end{split}
\label{np_strategy}
\end{align}
where $\mathbf{f}^p_{k-2},\mathbf{g}^p_{k-2}$ are concatenations of pure strategies to stage $(k-1)$. We denote the corresponding auxiliary matrix as $\tilde{Q}(s_{kl}) \in \tilde{\mathbb{Q}}_k$ for cyber state $\delta_l$, the one shot game value based on payoff matrix $\tilde{Q}(s_{kl})$ as $\tilde{v}_{k}(s_{kl})$, i.e.,
\begin{align}
\begin{split}
&\tilde{Q}(s_{kl}) \\
=&\tilde{r}(s_{kl})+\sum_{\delta_h\in S} \tilde{P}( s_{(k+1)h} |s_{kl})\cdot \tilde{v}_{k+1}(s_{(k+1)h}).
\end{split}
\label{tilde_Q}
\end{align}
Here each possible hybrid state $s_{kl}$ for time instant $k$ is calculated from a none pure strategy defined as~\eqref{np_strategy}. Similarly, the value is defined as
\begin{align}
\tilde{v}_{k}(s_{kl})=\max_{\tilde{Q}(s_{kl})\in \mathbb{\tilde{Q}}_k}v^*[\tilde{Q}(s_{kl})].
\label{tilde_v}
\end{align}
The following theorem shows that at every stage $k$, $\bar{v}_{k}^{p}(s_{kl})$ is greater than or equal to $\tilde{v}_{k}(s_{kl})$.
\\
\begin{thm}
Consider the value iteration for stage $k$ as a one shot robust game. %given a pure strategy history $h^{p}_{k}$, 
Based on $\bar{v}_{k}^{p}(s_{kl}) \geq 0$ of previous iteration, 
we define the robust game value  obtained at $k$ as~\eqref{pickv}. Then for $k=2,\cdots, K$, $\tilde{v}_{k}(s_{kl})$~\eqref{tilde_v} is upper bounded by $\bar{v}_{k}^{p}(s_{kl})$, i.e., $\tilde{v}_{k}(s_{kl}) \leqslant \bar{v}_{k}^{p}(s_{kl}).$
\label{robust}
\end{thm}
\begin{pf}
Since $\bar{v}_{k+1}^{p}(s_{(k+1)h})$ is a nonnegative scalar value, the extreme points of the set $\mathbb{Q}_{k}$ is a subset of the extreme points  of set $\mathbb{Q}^p_k$. Hence, by considering the value of matrix game $Q^{p} (s_{kl})\in \mathbb{Q}^p_k$ defined in~\eqref{eq:Q}, we will get the upper bound of the maximum game value from extreme points of $\mathbb{Q}_k$. 

Consider the following optimization problem for the system with constraint inequality~\eqref{optr} for any possible attacker's strategy vector $\mathbf{f}$ at each stage $k$
%From the system's perspective, to compute the value of the matrix game is equivalent with:
%\vspace{-5pt}
\begin{align}
%\begin{split}
%\vspace{-5pt}
\min_{\mathbf{g}} \quad & z\label{ob_z}\\
\text{subject to}\quad &z \geq \max_{\tilde{Q}(s_{kl})\in \tilde{\mathbb{Q}}_{k}}\mathbf{f}^{T} [\tilde{Q}(s_{kl})]\mathbf{g}.
%\end{split}
\label{optr}
\end{align}
As proven by~Lemma 5 in \cite{RGT},~\eqref{optr} is equivalent to the following constraint that considers only the extreme points
\begin{align}
\quad z \geq \max_{Q^{p}(s_{kl})\in \mathbb{Q}_{k}^p}\mathbf{f}^{T} [Q^{p}(s_{kl})]\mathbf{g},
\end{align}
For the worst-case $f$, the above is also true. Hence, let 
\begin{align}
\label{eq1}
v^p_{k}(s_{kl})
%=\max_{Q_{k}^{p}(h^{p}_{k},s_{l})\in \mathbb{Q}_{kp}}v^*[Q_{k}^{p}(h^{p}_{k},s_{l})]
=\max_{Q^{p}(s_{kl})\in \mathbb{Q}_k^p} \min\limits_{\mathbf{g}}\max\limits_{\mathbf{f}}\mathbf{f}^T[Q^{p}(s_{kl})]\mathbf{g}.
%\label{pickv}
\end{align}
For optimal policies $\mathbf{f}^{*}(s_{kl})$ and $\mathbf{g}^{*} (s_{kl})$, the above optimization problem~\eqref{eq1} results in a cost
\\\centerline{$
%\begin{align*}
 \max\limits_{Q^{p}(s_{kl})\in \mathbb{Q}_{k}^p}v^*[Q^{p}(s_{kl})].
%\end{align*} 
$}
However, $(\mathbf{f}^{*}(s_{kl}),\mathbf{g}^{*}(s_{kl}))$ can be non-pure strategies, meaning that when we apply  $(\mathbf{f}^{*}(s_{kl}),\mathbf{g}^{*}(s_{kl}))$  to calculate system dynamics such as equations~\eqref{dynamicgame}, they will not result in any extreme point of set $\mathbb{Q}_{k+1}$. 

Now consider the final stage $K$, we have
\\\centerline{$
%\begin{align*}
Q^p(s_{Kl})=r^p(s_{Kl}), \tilde{Q}(s_{Kl})=\tilde{r}(s_{Kl}),
%\end{align*}
$}
and use the $Q^p(s_{Kl})$ and $\tilde{Q}(s_{Kl})$ in the above proof, value $\tilde{v}_k(s_{Kl})$ from $\tilde{Q}(s_{Kl})$ is smaller than $\bar{v}_k^p(s_{Kl})$ from the extreme points auxiliary matrix $Q^p(s_{Kl})$, i.e., for $K$, the following inequality holds
\\\centerline{$
%\begin{align*}
\tilde{v}_k(s_{Kl}) \leqslant \bar{v}_k^p(s_{Kl}).
%\end{align*}
$}
Then, by induction, with the value $\tilde{v}_{k+1}(s_{(k+1)h})$ of iteration for stage $k+1, 2\leqslant k \leqslant K-1$ satisfies %%%%%(from $\mathbb{Q}_{(k+1)p}$):
%\begin{align*}
\\\centerline{$\tilde{v}_{k+1}(s_{(k+1)h})\leqslant \bar{v}_{k+1}^{p}(s_{(k+1)h}),$} 
%\end{align*}
and nonnegative payoff and state transition probability $r_{k}^{ij}\geqslant 0$ and $\tilde{P}_{k}^{ij}\geqslant 0$, replacing $\tilde{v}_{k+1}(s_{(k+1)h})$ by $v_{k+1}^{p}(s_{(k+1)h})$ in~\eqref{eq:Q} will make every entry of matrix $\tilde{Q}(s_{kl})$ smaller than matrix $Q^{p}(s_{kl})$. 
% since $r_{k}^{ij}\geqslant 0$ and $\tilde{P}_{k}^{ij}\geqslant 0$. %%% by definition. 
%The system is a minimizer, and is possible to get a smaller value at time k, with some %optimal strategy history $h^{*}_{k} \neq h^{p}_{k}$, 
With a similar argument in the next iteration for stage $k-1$, we have 
%\begin{align*}
\\\centerline{$
\tilde{v}_{k}(s_{kl}) \leqslant \bar{v}_{k}^{p}(s_{kl}).
$}
%\end{align*} 
%(Note that here $v_{k}^{s_{l}}(h^{p}_{k-1})$ is not the game value at $k$ either, because the strategies from 1 to $k-2$ must be pure to get it).
\end{pf}

Based on the above observation, we arrive at the suboptimal algorithm to compute the equilibrium solutions, illustrated in the Algorithm~\ref{finite}. Note that for keeping the physical state $x_{[k-T,k]}$ of the first stage of the game starts at $\hat{x}_0$, in the above Algorithm~\ref{finite} the $K$-stage game  starts at $k=T$. This does not affect our proofs in this section for considering $k=1,\dots, T$.
According to~Theorem~\ref{robust}, we use Algorithm~\ref{finite} to compute an upper bound of the value and the corresponding suboptimal strategy for every step. The % Nash Equilibrium selection 
function $\pi$ computes the strategy and robust value as defined in~\eqref{pickv}. 

The values of the finite stage game $\tilde{v}_k(s_{kl})$ and $\bar{v}_k^p(s_{kl})$ resulting from two auxiliary matrices $\tilde{Q}(s_{kl})$ $Q^p(s_{kl})$ are based on strategy concatenations that only differ at stage $k-1$ (i.e., the same and pure strategies from stages 1 to $(k-2)$). By value iteration backward to stage $1$, we compare the game value for all possible strategies and the robust game value $\bar{v}_1^p(s_{1l})$ of~Algorithm~\ref{finite} in the following theorem.
\begin{cor}
%Moreover, by following this iteration method,
~Algorithm~\ref{finite} yields an upper bound $v_1(s_{1l})$ for the value of the $K$-stage game, together with suboptimal strategies $\mathbf{f}_{a}$ and $\mathbf{g}_{a}$. 
\end{cor}
%\begin{pf}
The strategies $\mathbf{f}_{a}, \mathbf{g}_{a}$ of~Algorithm~\ref{finite} are possibly not pure. %strategy history $h^{p}_{K} \in H^{p}_{K}$. 
According to~Theorem~\ref{robust}, we obtain $\tilde{v}_{k}(s_{kl})\leqslant \bar{v}^{p}_{k}(s_{kl}),$ and the proof holds for every $k=2,\cdots,K$. Consider the value iteration for $k=1$, with 
$\tilde{v}_{2}(s_{2l})\leqslant \bar{v}^p_k(s_{2l})$, and $\text{\ }\quad Q^{ij}( s_{1l})$
$=r^{ij}(s_{1l})+\sum\limits_{\delta_h\in S}{P}^{ij}(s_{2h}|s_{1l})v^{p}_{2}(s_{2h}) \leqslant Q^{p,ij}(s_{2l}),$
%\end{align*}
%\normalsize
thus the true value of the K-stage game $v^{*}[Q(s_{1l})] \leqslant \bar{v}^{p}_{1}(s_{1l})$. The iterative value based on pure strategy auxiliary matrix sets $\mathbb{Q}_{k}^p, k=1,\cdots, K,$ obtained from~Algorithm~\ref{finite} is an upper bound for the game value.
%\end{pf}
Let $v^*[Q(s_{na})]$ represent the minimum total payoff of the system when the strategy is calculated given that there is no attack at all in $K$ stages, then $\bar{v}^{p}_{1}(s_{1l})-v^{*}[Q(s_{1l})] \leqslant \bar{v}^{p}_{1}(s_{1l})-v^*[Q(s_{na})]$, since  $v^*[Q(s_{na})] \leqslant v^{*}[Q(s_{1l})]$ when the system operates in normal state without sacrificing any control cost to play against attacks. The sub-optimality of value $\bar{v}^{p}_{1}(s_{1l})$ calculated from Algorithm 1 is then bounded though we do not know the true value $v^{*}[Q(s_{1l})]$ of the game.

\section{A Moving-Horizon Approach for Hybrid Stochastic Game}
\label{sec:algorithm}%Unlike the receding-horizon Stackelberg game with the leader and follower for correlated jamming attacks in~\cite{MartZhu_game}, the equilibrium considered in this section is different and the game state is hybrid.a time window of size $T$ is used, with physical state $x_{[k-T, k]}$ information by looking back $T$ stages and its associated cyber state $\delta_l$
In this section, we propose a moving-horizon algorithm to compute the saddle-point equilibrium strategy at each stage of the hybrid stochastic game. A saddle-point equilibrium strategy is computed at each stage $k$ by predicting anticipated future cost based on the hybrid state of the system $(x_{[k-T, k]},\delta_l )$. We develop Algorithm~\ref{finite_new} based on this concept, provides a scalable and a computationally tractable process, and compare the computational costs with Algorithm~\ref{finite}. The saddle-point equilibrium strategy and the value of the moving-horizon game at each stage involves solving finite zero-sum matrix games. By looking one stage ahead of the game state at $k$, predicting the physical dynamics $\mathbf{x}_{k+1}$ given any action pair, we obtain an objective function that reflects the payoff of the current stage and future expectation for computing the strategies at $k$. %The moving horizon process is illustrated as Figure~\ref{sg}. 

Given any action pair $(a_{ik},u_{jk})$ at stage $k$, we first update the state space form of the system dynamics $\mathbf{x}_{k+1}$ based on $\mathbf{x}_{[k-T,k]}$ as~\eqref{dynamicgame}. We view $\mathbf{x}_{k+1}$ as a function of $(\mathbf{x}_{[k-T, k]}, a_{ik},u_{jk})$, the immediate payoff function ${r}^{ij}(s_{(k+1)h})$ (for stage $k+1$) defined as~\eqref{payoff} is also a function of the current game state and players' actions. We denote this relation as $r_{k+1}(\mathbf{x}_{[k-T, k]}, a_{ik},u_{jk}, \delta_h)$ in the following algorithms to distinguish it between definition~\eqref{payoff}, where the latter is the payoff results from the action of two players' at stage $k+1$. Then, we compute the value of the matrix game at stage $k+1$, by looking one stage ahead and consider stage $k+1$ as the terminal stage of the game, the value of game stage $k+1$ is now directly calculated via for $r^{ij}(\mathbf{x}_{[k-T,k]},a_{ik},u_{jk},\delta_h)$, $h=1,2,3,$ $i\in \{1,\cdots, M\}, j \in \{1,\cdots, N\}$ as~\eqref{prev}:
\begin{align}
v^{ij}_{k+1}(x_{[x-T,k]},\delta_h)= \min\limits_{\mathbf{g}}\max\limits_{\mathbf{f}}(r(\mathbf{x}_{[k-T,k]},a_{ik},u_{jk},\delta_h)),
\label{prev}
\end{align}
where $v_{k+1}(x_{[k-T,k]},\delta_h) \in \mathbb{R}^{M \times N}$ is the value matrix of stage $k+1$ estimated at stage $k$ based on the current game state and all possible action pairs. With the predicted value from the next stage, define the moving-horizon auxiliary matrix for stage $k$ as:
%\footnotesize
\begin{align}
\begin{split}
&Q_{k}(s_{kl})\\=&r(s_{kl})
                         +\sum_{s_h\in S} P_k( s_{(k+1)h} |s_{kl})\cdot v_{k+1}(x_{[k-T,k]},\delta_h),
\end{split}
\label{Q_k}
\end{align}
The dot products of matrices ${P}_{k}( s_{(k+1)h} |s_{kl})$, $v_{k+1}(x_{[k-T,k]}$, $\delta_h)$ is an element-wise product of two elements at the same position of the two matrices. The value and stationary equilibrium strategies that Algorithm~\ref{finite_new} calculates at each stage $k$ is defined as following.
\begin{defn}
Given $s_{kl}$, $v_{k+1}(x_{[k-T,k]},\delta_h)$ as~\eqref{prev}, and auxiliary matrix $Q_{k}(s_{kl})$ as~\eqref{Q_k}, the value and equilibrium strategies at $k$ are defined as the following equation: 
\begin{align}
v(s_{kl})= \min\limits_{\mathbf{g}_k(s_{kl})}\max\limits_{\mathbf{f}_k(s_{kl})} \mathbf{f}_k(s_{kl})^T Q_{k}(s_{kl}) \mathbf{g}_k(s_{kl}),
\label{v_k}
\end{align}
where we treat the auxiliary matrix $Q_{k}(s_{kl})$ as the payoff matrix of a zero-sum game of stage $k$. 
%The value and equilibrium strategies of the matrix game $Q_{k}(s_{kl})$ consist with the meaning of value and equilibrium strategies in Definition~\ref{zero_sum}.
\end{defn}

At each stage $k$, we repeat calculating $Q_{k}(s_{kl})$ and the corresponding value and equilibrium strategies, then update the system dynamics by the strategies for computation of next stage. The complete process is summarized as Algorithm~\ref{finite_new}. 
\begin{alg}
%\label{finite_new}
\textbf{: Moving-Horizon Algorithm for A Hybrid Stochastic Game}\\
%\begin{algorithmic}
\textbf{Input}: System model parameters and game parameters.
\\\textbf{Initialization}: $\hat{\mathbf{x}}_{[0,T]}$.
\\\textbf{Iteration}: For $k=T, \cdots, K+T-1$,                                                                                 
 $s_{kl}=(x_{[k-T,k]},\delta_l),$  $l=1,2,3$:  
get  the auxiliary matrix~\eqref{Q_k};
%\[Q_{k}(s_{kl}) = r_{k}(s_{kl})+\sum_{s_h\in S} \tilde{P}_{k}( s_{(k+1)h} |s_{kl})\cdot v_{k+1}(x_{[k-T,k]},\delta_h),\]
compute the value and equilibrium strategies of every matrix game:\\
$v(s_{kl})= \min\limits_{\mathbf{g}(s_{kl})}\max\limits_{\mathbf{f}(s_{kl})} \mathbf{f}(s_{kl})^T Q_{k}(s_{kl}) \mathbf{g}(s_{kl})$,\\
$\mathbf{f}_k^{*}(s_{kl})=\arg \max\limits_{\mathbf{f}_k(s_{kl})}\mathbf{f}_k(s_{kl})^T Q_{k}(s_{kl} )\mathbf{g}_k^*(s_{kl})$, \\
$\mathbf{g}_k^{*}(s_{kl})=\arg \min\limits_{\mathbf{g}_k(s_{kl})} [\mathbf{f}_k^{*}(s_{kl})]^T Q_{k}(s_{kl}) \mathbf{g}_k(s_{kl})$.\\
Update the system dynamics with strategies $\mathbf{f}_k^{*}(s_{kl}),\mathbf{g}_k^{*}(s_{kl}),$ $l=1,2,3$ as described in~\ref{dynamicgame} for the next stage.
\\\textbf{Return}: the concatenation of strategies for both players $\mathbf{f}=\{f_k^{*}(s_{kl})\},\mathbf{g}=\{\mathbf{g}^*_k(s_{kl})\}$ and the value sequence $v_{k}(s_{kl}),k=T,\cdots, K+T, l=1,2,3$.
\label{finite_new}
\end{alg}
To get the total payoff till stage $k$ by Algorithm~\ref{finite_new}, we plug the strategies $\mathbf{f}, \mathbf{g}$ into the system dynamics and calculate the sum of payoff for all stages. It is worth noting that~Algorithm~\ref{finite_new} reduces the computational overhead for the hybrid stochastic game. The complexity of Algorithm~\ref{finite_new} is equivalent to the complexity of solving $(KMN)$ times of minimax problem with an $M \times N$ payoff matrix, while the complexity of suboptimal Algorithm~\ref{finite} is equivalent to the complexity of solving $((MN)^{K})$ times of minimax problem with an $M\times N$ payoff matrix. %Because it takes the total expected payoff of $K$ stages ahead as an objective function. %Numerical comparisons are shown in Section~\ref{sec:simulation}. %This is because Algorithm~\ref{finite} looks $K$ stages ahead at once and compute a robust game for every iteration. The advantage of suboptimal Algorithm~\ref{finite} is to provide an upper bound of the total finite cost..

\begin{rem}
%Given the sets of models for each component of the subsystems and attacks, 
The system dynamics are defined by a sequence of action pairs $(a_{ik},u_{jk})$ randomly chosen by the attacker and the system, and are equivalent with a system that randomly switches among $N$ subsystems according to the stochastic game strategies $\mathbf{f}_k(s_{kl})$. The strategy sequences $\mathbf{f}_k^*(s_{kl})$, $\mathbf{g}_k^*(s_{kl})$ of the stochastic game converge to $\mathfrak{f}^l,\mathfrak{g}^l $, $l=1,2,3$, i.e.,
\\\centerline{$\mathfrak{f}^l=\lim\limits_{k\to\infty}\mathbf{f}_k^*(s_{kl}), \mathfrak{g}^l=\lim\limits_{k\to\infty}\mathbf{g}_k^*(s_{kl}), l=1,2,3,$}
if updating system dynamics at stage $k+1$ by ($\mathfrak{f}^l,\mathfrak{g}^l$) results in:
\\\centerline{$\lim \limits_{k \to \infty} Q_{k}(s_{kl})=\lim \limits_{k \to \infty} Q_{k}(s_{(k+1)l}), l=1,2,3.$}
%\end{prop}
This is because according to Algorithm~\ref{finite_new}, $\mathbf{f}^*_k(s_{kl})$, $\mathbf{g}^*_k(s_{kl}),l=1,2,3$ are the saddle-point equilibrium strategies for the auxiliary matrices $Q_{k}(s_{kl}),l=1,2,3$. When the strategy sequences of both players converge, the switched system dynamics converge to a discrete-time Markov jump linear system (with delays when the attacker's strategies include replay attacks), and the stability properties of the system that switches among stable and unstable subsystems is analyzed by~\cite{delay_mlj} and~\cite{switch_unstable}.\end{rem} 
%It is possible that some subsystems $u_{jk}, j \in \{1,\cdots, N\}$ are unstable under specific types of attacks, and the system switches among stable and unstable subsystems. Stability properties of continuous time linear switched systems including unstable modes are analyzed by. To guarantee exponential stability, the total activation time of unstable subsystems need to be relatively small compared with that of stable subsystems. More analysis of system stability conditions based on the moving horizon stochastic game framework will be an avenue of future work. With the heuristic moving horizon algorithm, by calculating strategies of the stochastic game for a large enough stage number, we obtain the switched dynamic process of the system under different types of attacks, and check whether stability conditions are violated

\section{Comparison of Algorithms}
\label{sec:simulation}
One advantage of the moving horizon Algorithm~\ref{finite_new} is its faster computation speed. Table~\ref{alg_compare} shows Matlab simulation time for different $K$-stage games, all with the same size of action space for the system and attacker. When $K$ increases, the difference between algorithm speed also increases.
We compare the cost of the strategies provided by the suboptimal Algorithm~\ref{finite} and~Algorithm~\ref{finite_new}.
The example studied is an unstable batch reactor, a four dimensional system~(see \cite{ncs}, Section IV.A for model parameters).

We first show the case under replay attacks, when the system is equipped with two controllers, one steady state Kalman filter, and the corresponding $\chi^2$ detector. An optimal LQG controller $u^*_K$ is denoted as controller $1$, and a non-optimal controller $(u^{*}_{k} + \Delta u_{k})$~(\cite{replay}) with higher replay detection rate as controller $2$. System's action space includes: subsystem $u_{1k}$ with controller $1$ and subsystem  $u_{2k}$ with controller $2$. For illustration, we show the case when the attacker's action space are discretized replay attack time window size $\{10s, 20s, 30s, 40s\}$ in simulation. We design switched control policy for the system under replay attacks with initial mode $\delta_{2}$, (i.e.,~$p(\delta^{1}_{2})=1$), we compare the system's strategies and total payoff when applying suboptimal strategies of Algorithm~\ref{finite} and real-time receding horizon Algorithm~\ref{finite_new} in a finite game of stage $K=50$.

\begin{figure}[b!]
%\vspace{-5pt}
\centering
\includegraphics [width=0.42\textwidth]{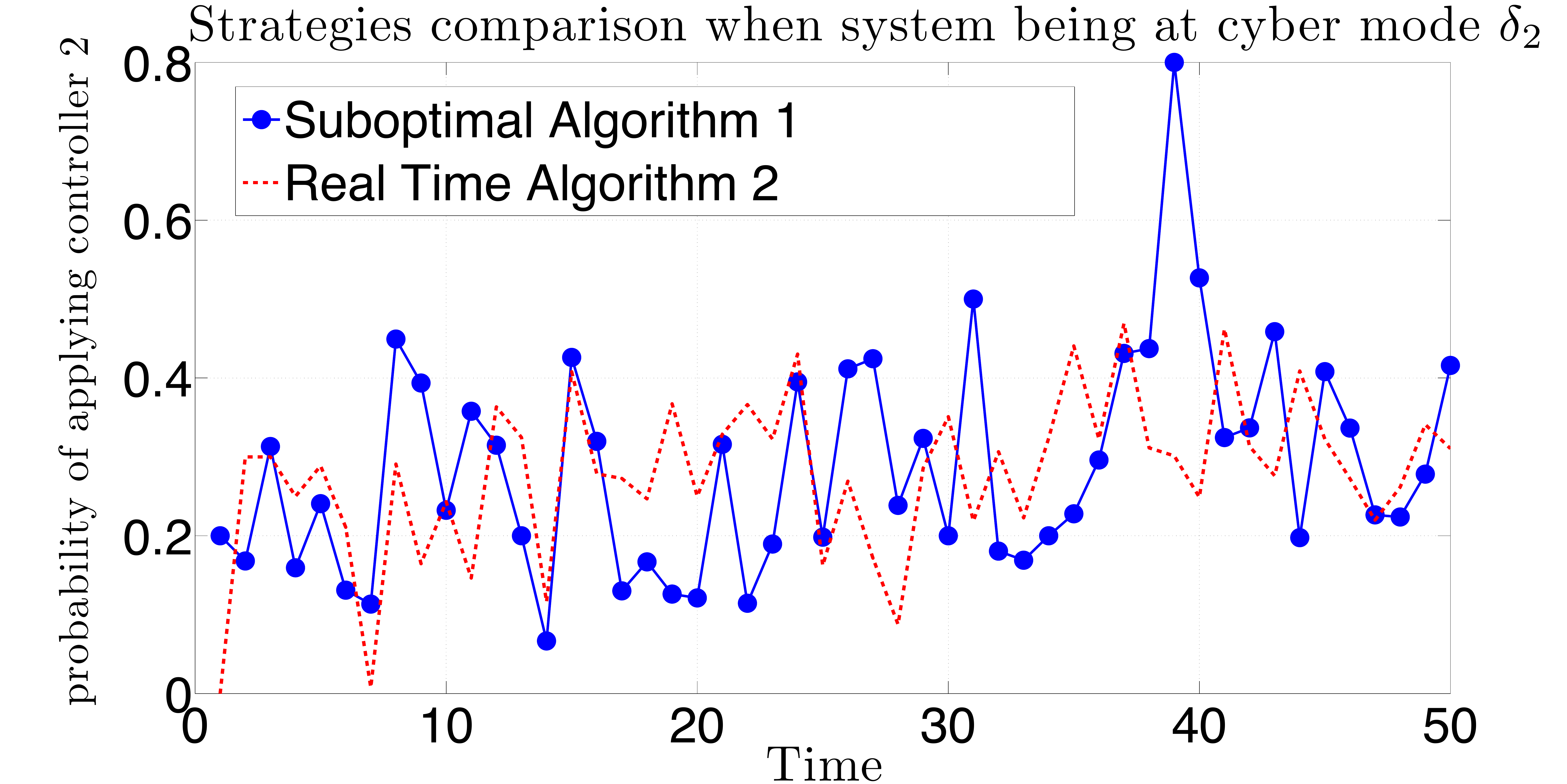}
\vspace{-8pt}
\caption{Strategies comparison of two algorithms for system under replay attack--the probability of switching to subsystem 2 at mode $\delta_{2}$ of every $k$. }
\label{ads}
\vspace{-5pt}
\end{figure}

\begin{figure}[b!]
\vspace{-5pt}
\centering
\includegraphics [width=0.42\textwidth]{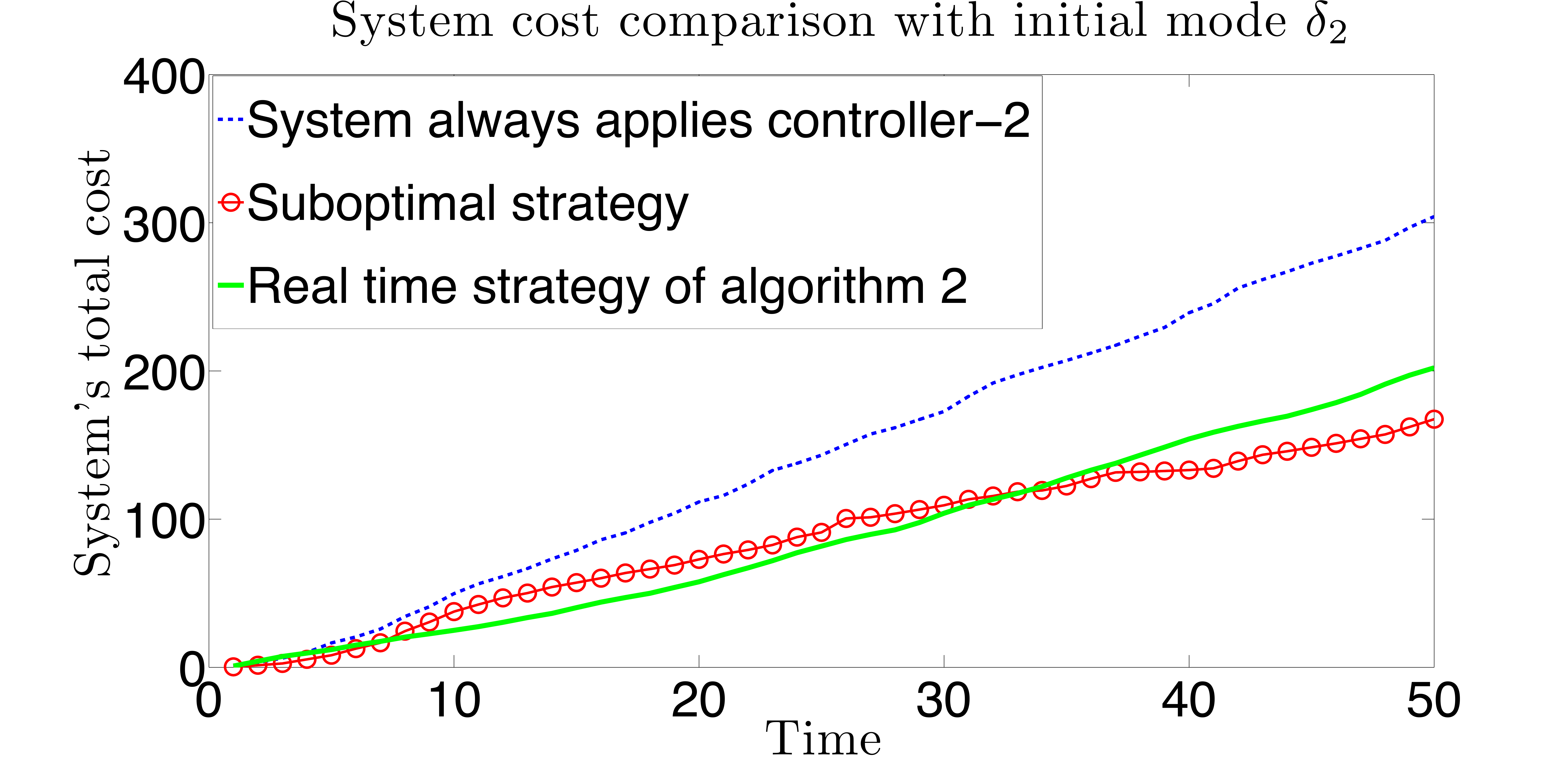}
\vspace{-15pt}
\caption{Cost comparison of system applying different strategies at mode $\delta_{2}$. Applying the suboptimal strategy provides the smallest cost, and the strategy of the real time algorithm is better than the one of a non-game approach.}
\label{cost2_c2}
\end{figure}

\begin{figure}[t!]
%\vspace{-5pt}
\centering
\includegraphics [width=0.42\textwidth]{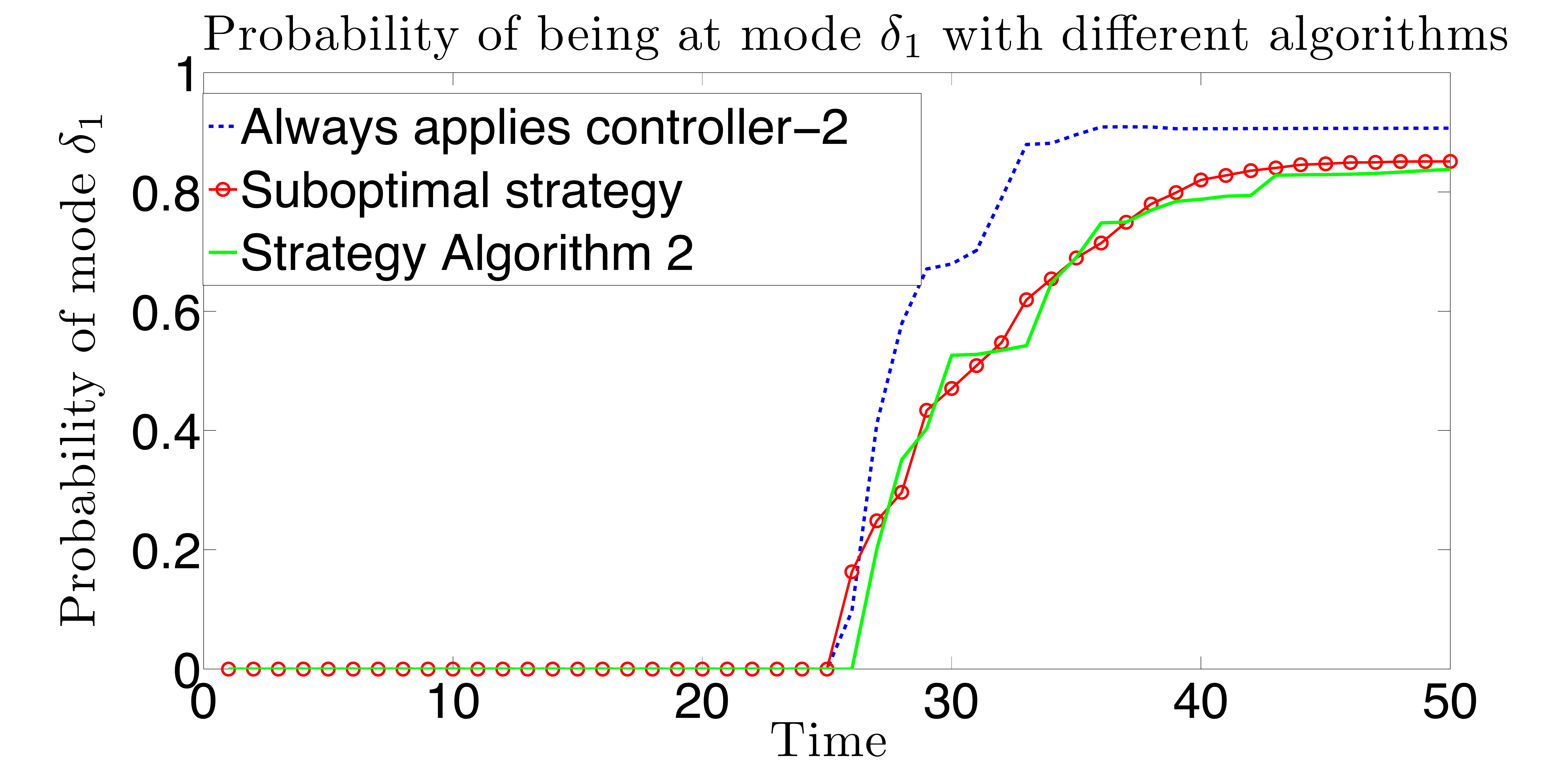}
\vspace{-10pt}
\caption{Comparison of the probability of the system being at mode $\delta_{1}$ for different strategies. 
Game strategies provide similar detection rate with the non-switching policy.}
\label{ps1_c1}
\vspace{-5pt}
\end{figure}

Figure~\ref{ads} shows the probability of switching to Controller~$2$ at every stage according to different algorithms.
Three cases are shown in Figure~\ref{cost2_c2}--when the system applies the strategy of Algorithm~\ref{finite}, the strategy of Algorithm~\ref{finite_new}, and only the subsystem $2$ with higher replay detection rate through all stages. 
Figure~\ref{ps1_c1} shows the probability that system being at mode $\delta_{1}$ (successfully detected an attack), when applying strategies obtained from the two algorithms and always choosing subsystem $2$. Applying a game strategy, randomly switching between subsystems results in a lower cost, while not sacrificing the detection rate significantly. 

\begin{table}%[t]
\centering
\begin{tabular}{|c|c|c|}
  \hline
   K  & \iffalse Time of \fi real time algorithm & \iffalse Time of \fi suboptimal algorithm \\ \hline
   20 &   1.8054s    &  6.7346s    \\  \hline
   50 &   4.9968s    &  58.6144s  \\ \hline  
   100&  8.3827s    &  2073.2928s   \\ \hline
    500&  41.0342s  & 20h    \\ \hline
\end{tabular}
\caption{Elapsed time comparison of two algorithms} %for different total stage numbers}
\label{alg_compare}
\end{table}

\begin{figure}[t!]
\centering
\includegraphics [width=0.42\textwidth]{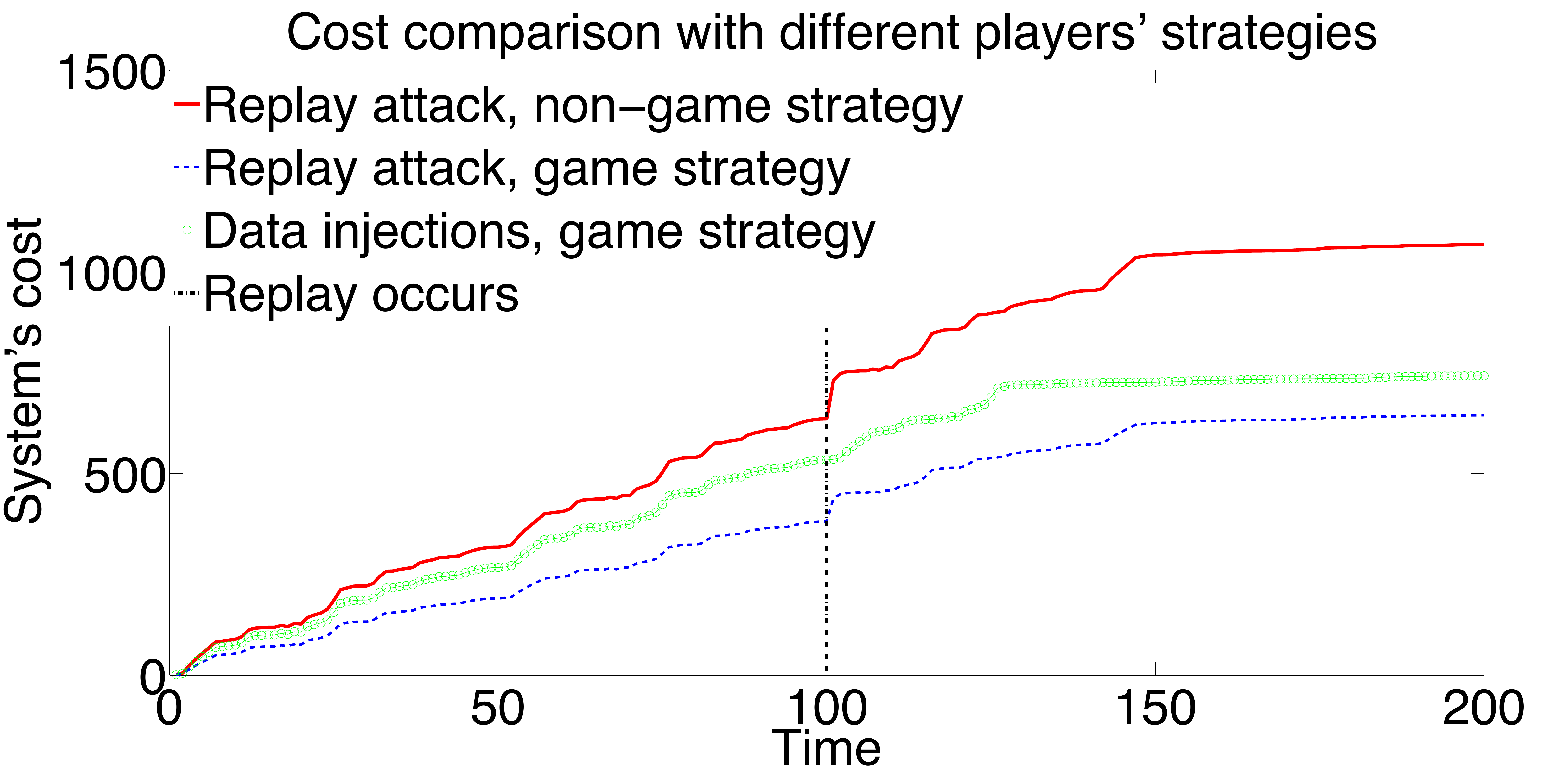}
\vspace{-15pt}
\caption{Cost comparison when system and the attacker apply different strategies. The replay attack occurs at $T_2=100s$.}
\label{long_time}
\vspace{-5pt}
\end{figure}

For game strategies designed for multiple types of attack, Figure~\ref{long_time} shows the case when attacks are successfully detected and the system reaches the cyber mode $\delta_1=safe$, the quadratic cost of the system converge. When replay finally occurs at $T_2=100s$, with a game-theoretic strategy, the cost of the system is smaller than the cost when system always applies a controller with higher cost and higher detection rate. Data injection attacks shown in Figure~\ref{long_time} appear during $k=30,31,\dots,50$.   

These figures illustrate that the real-time strategy results a higher cost than the suboptimal system strategy, and they both provide lower control costs compared to the non-game-theoretic approach. The non-game-theoretic approach provides only a slightly higher probability of being at the safe mode in $K$ stages. By introducing the game strategy, i.e., switching between multiple subsystems, we do not sacrifice the payoff of the system while providing an acceptable detection rate, even we discretize the attacker's action space in the game framework. For instance, Figures~\ref{cost2_c2} and~\ref{ps1_c1} show the result when the actual replay attack occurs at $T_2=25s$, and the game strategies are calculated with action space $A_t=\{10s,20s,30s,40s\}$. Since $25$ is in between $[20,30]$, and the error of the action space discretization is bounded, a game strategy calculated via finite action space improves the system's performance.

\iffalse
The linearized model parameters are
\footnotesize
%\begin{align*}
\centerline{$
\mathbf{A}&=\begin{bmatrix}1.38 &   -0.2077 &  6.715 &  -5.676\\
   -0.5814 &-4.29   &    0  &    0.675\\
   1.067 &  4.273 &    -6.654&  5.893\\
   0.048  & 4.273 &    1.343 &  -2.104 \end{bmatrix},$}\\
\centerline{$
\mathbf{B}&=\begin{bmatrix}0 & 0\\
   5.679 & 0\\
   1.136 & -3.14\\
   1.136 & 0\end{bmatrix}, \quad \mathbf{C}=\begin{bmatrix}1 &  0 &  1 &  -1\\ 0&  1 & 0 &  0\end{bmatrix},\quad \mathbf{D}=\mathbf{0}.
   $}
%\end{align*}
\normalsize
\fi

\section{Conclusion}
\label{sec:concl}
In this work, we have proposed a zero-sum hybrid stochastic game model to capture the interactions between a cyber-physical system and an attacker --- switching policy for the system under different types of sensor attacks. This framework allows us to find a control policy by calculating stationary strategy of the game with information of the system's physical dynamics and cyber modes. We design a suboptimal value iteration algorithm for a finite horizon game, which considers a saddle-point equilibrium of a robust stochastic game at each iteration. To reduce the computational complexity, a real-time moving-horizon algorithm is then developed. Based on the concept of saddle-point equilibrium for the hybrid stochastic game, at each stage, we look one stage ahead to calculate anticipated future value. The stability conditions of the system under multiple types of attacks based on the stochastic game framework, and an alternative algorithm with unknown transition matrix or payoffs will be our future work.

\bibliographystyle{agsm}        
{  \small 
\bibliography{gamejs}
}

\end{document}